\documentclass[preprint,12pt]{elsarticle}

\usepackage{graphicx}
\usepackage{comment}
\usepackage{amssymb}

\usepackage{lineno}
\usepackage{lineno,hyperref}
\usepackage{amsmath,amsfonts,amsthm,amssymb,graphicx,subfigure,mathtools,tikz,hyperref,bm,blindtext,float}
\usepackage{diagbox}
\usepackage{geometry}
\usepackage{natbib}
\usepackage{indentfirst}
\usepackage{subfigure}
\usepackage{array}

\usepackage{multirow}
\usepackage{bm}
\usepackage{longtable}

\journal{Journal Name}

\begin{document}

\begin{frontmatter}


\title{Physics-informed semantic inpainting: Application to geostatistical modeling}



\author[zju,brown]{Qiang Zheng}
\author[zju]{Lingzao Zeng\corref{cor2}}
\cortext[cor2]{Corresponding author}
\ead[cor2]{lingzao@zju.edu.cn}
\author[sdu]{Zhendan Cao\corref{cor3}}
\author[brown,PNNL]{George Em Karniadakis\corref{cor1}}

\address[zju]{Institute of Soil and Water Resource and Environmental Science, Zhejiang Provincial Key Laboratory of Agricultural Resources and Environment, College of Environmental and Resource Sciences, Zhejiang University, Hangzhou, China}
\address[brown]{Division of Applied Mathematics, Brown University, Providence, RI, USA}
\address[sdu]{Department of Geology and Geological Engineering, South Dakota School of Mines and Technology, Rapid City, SD, USA}
\address[PNNL]{Pacific Northwest National Laboratory, Richland, WA, USA}

\begin{abstract}
A fundamental problem in geostatistical modeling is to infer the heterogeneous geological field based on limited measurements and some prior spatial statistics. Semantic inpainting, a technique for image processing using deep generative models, has been recently applied for this purpose, demonstrating its effectiveness in dealing with complex spatial patterns. However, the original semantic inpainting framework incorporates only information from direct measurements, while in geostatistics indirect measurements are often plentiful. To overcome this limitation, here we propose a {\em physics-informed} semantic inpainting framework, employing the Wasserstein Generative Adversarial Network with Gradient Penalty (WGAN-GP) and jointly incorporating the direct and indirect measurements by exploiting the underlying physical laws. Our simulation results for a high-dimensional problem with 512 dimensions show that in the new method, the physical conservation laws are satisfied and contribute in enhancing the inpainting performance compared to using only the direct measurements.
\end{abstract}




\end{frontmatter}


\section*{1. Introduction}
In recent years, deep learning has achieved great success in image processing, and it has been introduced to tackle geophysical problems, such as extreme weather detection \cite{liu2016application,racah2017extremeweather}, land cover and land use classification \cite{zhang2019remote,helber2019eurosat}, which cannot be adequately handled by traditional methods. In geological modeling, one common problem is to delineate the field heterogeneity from sparse measurements, i.e., inferring the entire field based on several point measurements and prior spatial statistics. This task is very similar to the concept of {\em semantic inpainting} originated from deep learning realm, which takes advantage of deep generative models for image reconstruction based on some known pixels~\cite{yeh2017semantic}.

Among different deep generative models, generative adversarial networks (GANs) \cite{goodfellow2014generative} are particularly suited for geophysical applications as they can model high-dimensional, multi-modal distributions. The "magic" of GANs in cross-domain image translation has been applied to build bidirectional mappings between the parameter and state fields in subsurface models \cite{sun2018discovering}. GANs have also been employed to define a low-dimensional parameterization for complex geological fields, hence allowing for a probabilistic inversion using Bayesian methods, like Markov Chain Monte Carlo (MCMC)~\cite{laloy2018training}. The low-dimensional embedding widely exists in the physical world, and it is beneficial to gain insights by finding the embedding and the corresponding mapping to the real process. In the semantic inpainting, the discovery of underlying embedding is certainly of great importance. Therefore, as state-of-the-art generative models, GANs are greatly qualified for this task.

Recently, there have been some works that combine semantic inpainting with geostatistical modeling using GANs, such as the modeling of fluvial patterns \cite{dupont2018generating} and realistic 3D reservoir facies \cite{zhang2019generating}. Specifically, they trained a GAN model using unconditional realizations to learn a mapping from a latent simple distribution, e.g. Gaussian, to a real data distribution and then found the optimal latent variable underlying the known measurements, and eventually reconstructed the incomplete image data based on the optimal latent variable using the pre-trained GAN model. However, the existing studies only considered direct measurements for semantic inpainting, while in many geological scenarios, indirect measurements are more abundant than the direct ones. For example, the statistical modeling of heterogeneous hydraulic conductivity field is a common problem in hydrogeology. Since the direct measurements are usually very expensive, it would be greatly beneficial to assimilate indirect measurements, e.g. pressure heads, which are relatively easier to obtain.

Inspired by the physics-informed neural networks (PINNs)~\cite{raissi2019physics}, and their applications \cite{raissi2019deep,tartakovsky2018learning} as well as the physics-informed GANs \cite{yang2018physics}, we aim to fuse the information from indirect measurements through adding physical constraints, i.e., the residual of the conservation law, to the loss function. In the PINN framework, usually a fully-connected network architecture is adopted and the differential terms are calculated directly by automatic differentiation \cite{baydin2018automatic}, since the inputs contain spatial coordinates; hence, no mesh generation is required. However, the semantic inpainting workflow takes images as input with no coordinates information, and employs convolutional network architectures to extract the spatial patterns. Therefore, automatic differentiation cannot be used in the workflow of semantic inpainting. Fortunately, the convolutional kernel is a smoothed finite difference operator, namely the Sobel filter~\cite{gao2010improved}, which can be employed to compute the differential terms in the partial differential equation (PDE). This technique has also been utilized to build physics-constrained surrogates for high-dimensional problems, and it was suggested that the Sobel filter is more efficient than automatic differentiation to obtain spatial gradients~\cite{zhu2019physics}.

In this work, we extend the original semantic inpainting framework~\cite{yeh2017semantic} to a physics-informed version, for the purpose of fusing information from both the direct and indirect measurements. Specifically, we propose a physics-informed Wasserstein GAN with gradient penalty (WGAN-GP) method for semantic inpainting, in which the Sobel filter is employed to construct the physical constraints, i.e, the PDE that expresses the conservation law. We then test its performance on a high-dimensional problem with 512 dimensions. The rest of the paper is organized as follows. In section 2, we introduce the physical model and present the inpainting task. In section 3, we propose the physics-informed WGAN-GP for semantic inpainting. Then, in section 4, we evaluate the proposed method using a high-dimensional case. The conclusions are given in the last section.

\section*{2. Problem Statement}
We consider steady-state flow in saturated media. The problem is to estimate the hydraulic conductivity field $K$ $\rm (L/T)$ from several direct measurements and a larger number of hydraulic head $h$ ($\rm L$) measurements. Here, we aim to achieve better estimation of the $K$ field by leveraging the $h$ measurements and the physical connection between them.

The hydraulic head ($h$) is connected with the hydraulic conductivity ($K$) via the following equation
\begin{equation}
    \nabla{\cdot [K(\textbf{x})\nabla{h(\textbf{x})}]} + q(\textbf{x}) =0,
    \hspace{1cm}
    \textbf{x} \in \mathcal{X}
    \label{eq:steady-state flow}
\end{equation}
with boundary conditions
\begin{equation}
    \begin{split}
        h(\textbf{x}) &= h_D(\textbf{x}),
        \hspace{1cm}
        \textbf{x} \in \Gamma_D,\\
        \nabla{h(\textbf{x})}\cdot n &= g(\textbf{x}),
        \hspace{1cm}
        \textbf{x} \in \Gamma_N,
    \end{split}
    \label{eq:boundary conditions}
\end{equation}
where $\textbf{x}$ is the coordinates and $q(\textbf{x})$ ($\rm L^3/T$) is the source (or sink) term; $n$ is the unit normal vector to the Neumann boundary $\Gamma_N$, and $\Gamma_D$ is the Dirichlet boundary. Given a number of two-dimensional $K$ realizations, we can obtain the corresponding $h$ realizations by using the numerical solver ModFlow \cite{harbaugh2005modflow}. Therefore, this equation can provide {\em paired} $K$ and $h$ datasets in the form of two-dimensional images for training the GAN models.

\section*{3. Methodology}
\subsection*{3.1 Physics-informed WGAN-GP}
The initial GAN proposed by~\cite{goodfellow2014generative} is composed of a generator and a discriminator, competing with each other in a zero-sum game to reach the Nash equilibrium, in which the generator produces fake samples very similar to the real ones. Let $\mathcal{G}$ represents the generator that builds a mapping for two distributions, i.e., $\mathcal{G}:P_\textbf{z} \to P_{\rm data}$; namely, it takes $\textbf{z} \in P_{\textbf{z}}$ as input and generates a fake sample $\mathcal{G}(\textbf{z})$ that is assumed to share the same support as the training dataset $\textbf{y} \in P_{\rm data}$. The discriminator $\mathcal{D}$ is used to detect whether its input is sampled from the real data distribution $P_{\rm data}$ or the fake data distribution $P_{\mathcal{G}(\textbf{z})}$. Therefore, the generator and discriminator, both neural networks, are placed in a competition, that is the generator tries to fool the discriminator, while the discriminator, with access to real data, tries to avoid being fooled. When the discriminator succeeds, it also provides useful feedback to help the generator improve and upon convergence both players have acquired the maximum knowledge given the real dataset.

In the original GANs, the neural networks are trained via the following loss function:
\begin{equation}
    \centering
    \min_{\mathcal{G}}\max_{\mathcal{D}}E_{\textbf{y}\sim P_{\rm data}}[log(\mathcal{D}(\textbf{y})]+E_{\textbf{z}\sim P_\textbf{z}}[log(1-\mathcal{D}(\mathcal{G}(\textbf{z})))].
\end{equation}
In practice, the generator and discriminator are trained iteratively, aiming to make $P_{\mathcal{G}(\textbf{z})}$ approach $P_{\rm data}$, or equivalently, minimize the
Jensen-Shannon (JS) divergence between them. However, the JS divergence cannot always provide effective gradients for the generator, especially when the two distributions concentrate on low dimensional manifolds~\cite{arjovsky2017wasserstein}. Consequently, the training of original GANs is usually unstable. To address this issue, Wassestein GANs (WGANs) were proposed by switching the JS divergence to the Wasserstein-1 distance, which is continuous and differentiable almost everywhere with respect to the parameters in the generator under a temperate constraint~\cite{arjovsky2017wasserstein}. To be more precise, the constraint to be imposed is Lipschitz continuity for the discriminator, which is realized by forcing the parameters of the discriminator to be bounded during training. To make the training of WGANs more stable, WGANs with gradient penalty (WGAN-GP) were developed to replace the weights clipping with adding gradient penalty to the the discriminator~\cite{gulrajani2017improved}. Specifically, the loss functions for the generator and discriminator of WGAN-GP are defined as
\begin{equation}
    \begin{split}
    \mathcal{L}_{\mathcal{G}}&=-E_{\textbf{z}\sim P_{\textbf{z}}}[\mathcal{D}(\mathcal{G}(\textbf{z}))],\\
     \mathcal{L}_{\mathcal{D}}&=E_{\textbf{z}\sim P_{\textbf{z}}}[\mathcal{D}(\mathcal{G}(\textbf{z}))]-E_{\textbf{y}\sim P_{\rm data}}[\mathcal{D}(\textbf{y})] + \lambda E_{\hat{\textbf{z}}\sim P_{\hat{\textbf{z}}}}[(\left\Vert \nabla_{\hat{\textbf{z}}}\mathcal{D}(\hat{\textbf{z}})\right\Vert_2 -1)^2],
    \end{split}
\label{eq:GAN loss}
\end{equation}
where $P_{\hat{\textbf{z}}}$ is the distribution generated by uniformly sampling along the straight lines between pairs of points sampled from $P_{\rm data}$ and $P_{\mathcal{G}(\textbf{z})}$, and $\lambda$ is the coefficient of the gradient penalty.

In this paper, we select the WGAN-GP with convolutional architectures as the base model. The key point in this work is to enable WGAN-GP to incorporate the information from physics, i.e. the existing PDEs, thus physical constraints should be added to the loss function when training WGAN-GP. In the networks based on convolutional architectures, nevertheless, automatic differentiation cannot be used for gradient calculation, because in convolutional networks the inputs are pixel values rather than the spatial coordinates. Here, we apply the Sobel filter to estimate the horizontal and vertical spatial gradients by imposing one convolution with the following $3\times3$ kernels, respectively:
\begin{equation}
\centering
     \mathcal{H} = \left[ \begin{array}{ccc}
        1 & 0 & -1 \\
        2 & 0 & -2 \\
        1 & 0 & -1 \end{array} \right],
\hspace{0.5cm}
\mathcal{V} =  \left[ \begin{array}{ccc}
        1 & 2 & 1 \\
        0 & 0 & 0 \\
        -1 & -2 & -1 \end{array} \right].
\label{sobel_filter}
\end{equation}
This method of calculating spatial gradients via convolutional kernel is equivalent to a smoothed finite difference method. Zhu et al. \cite{zhu2019physics} applied this method to physics-constrained surrogate modeling with deep learning models and suggested that Sobel filter is more efficient (although less accurate) than using automatic differentiation to obtain spatial gradients.

\begin{figure}
    \centering
    \hspace{0.7cm}
    \includegraphics[width=0.9\textwidth]{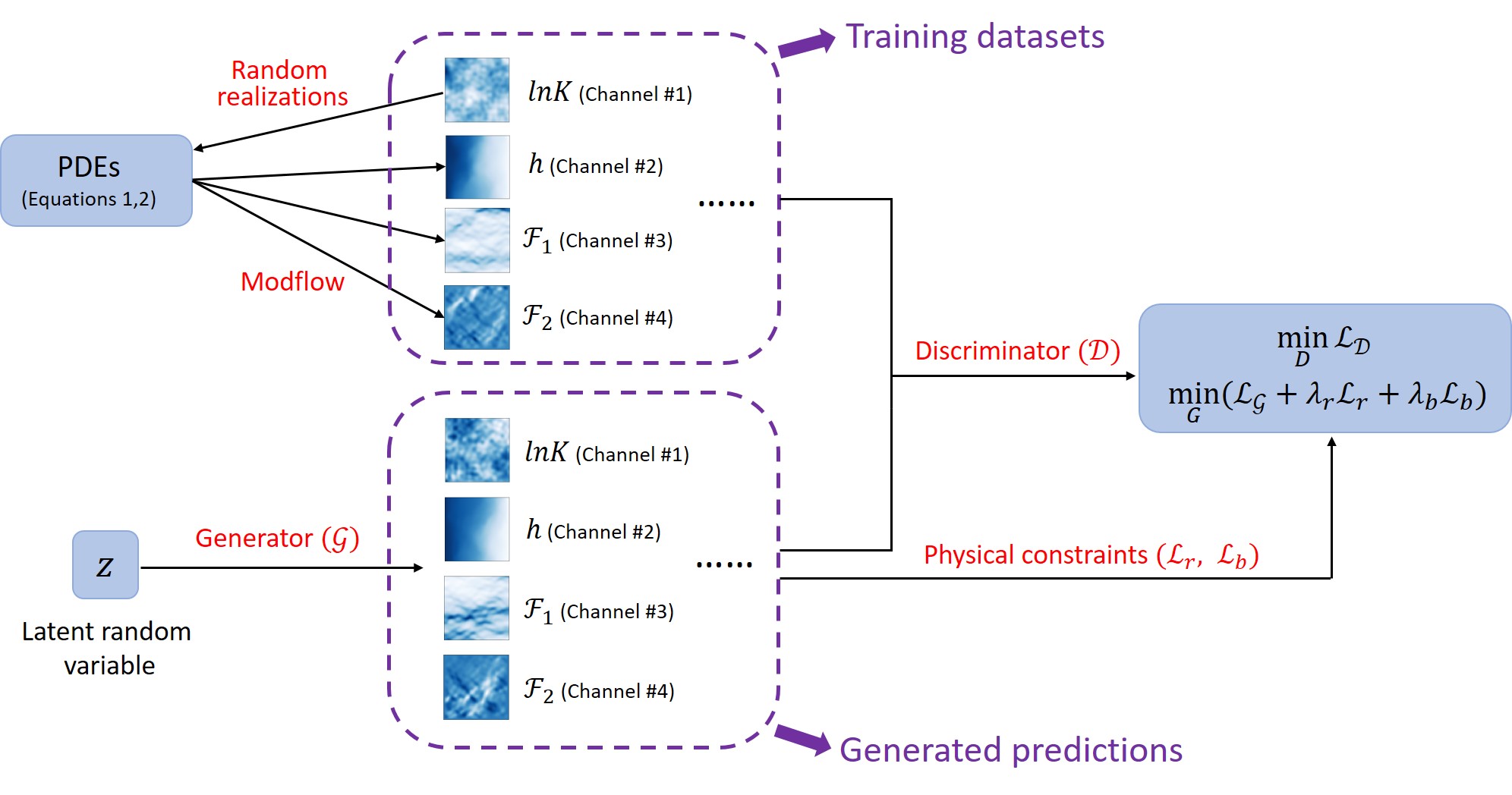}
    \caption{Schematic of pre-training the physics-informed WGAN-GP to be used in subsequent semantic inpainting. The WGAN-GP framework is based on the original version \cite{gulrajani2017improved}, containing one generator ($\mathcal{G}$) and one discriminator ($\mathcal{D}$), and both of them use convolutional architectures to cope with high-dimensional images. The datasets used for training are image-like data with four channels representing four physical variables, i.e., ${\rm ln}K$, $h$, $\mathcal{F}_1$ and $\mathcal{F}_2$. The random realizations of ${\rm ln}K$ generated by Karhunen-Lo\`eve expansion are fed to equations (\ref{eq:steady-state flow}) and (\ref{eq:boundary conditions}) to generate the corresponding numerical solutions ($h$, $\mathcal{F}_1$ and $\mathcal{F}_2$) by using ModFlow. The four channels of each sample in the training datasets are paired to satisfy the PDE. To help $\mathcal{G}$ incorporate the information from the PDE, the residual loss ($\mathcal{L}_r$, equation (\ref{eq:residual loss})) and boundary loss ($\mathcal{L}_b$, equation (\ref{eq:boundary loss})) are added to the original generator loss ($\mathcal{L}_{\mathcal{G}}$, equation (\ref{eq:GAN loss})). The $\mathcal{G}$ and $\mathcal{D}$ are trained iteratively following the basic training scheme of the original WGAN-GP. Upon convergence to the trained models, we can sample a new Gaussian noise $\textbf{z}$ and use $\mathcal{G}(\textbf{z})$ to obtain the generated predictions.}
    \label{fig:physics-based WGAN-GP}
\end{figure}

Considering that the PDE for saturated flow is second-order, which may bring some inaccuracy in the gradients calculation with the Sobel filter, we introduce an additional variable, namely the flux $\mathcal{F}$, to bypass calculating the second-order gradients. Therefore, the equation (\ref{eq:steady-state flow}) can be recast into a system of two equations, i.e.
\begin{equation}
    \begin{split}
        \mathcal{F}&= -K\nabla{h},\\
        \nabla{\cdot \mathcal{F}}&=q.
    \end{split}
    \label{eq: a system of equations}
\end{equation}
As shown in Figure \ref{fig:physics-based WGAN-GP}, the parameter ($K$), state ($h$) and additional variable ($\mathcal{F}$) are distributed as different channels with the same image sizes. It is worth noting that the flux $\mathcal{F}$ can be divided into horizontal and vertical ones, so they account for two channels. Therefore, based on equation (\ref{eq: a system of equations}), the residual of the PDE in the form of the loss can be written as
\begin{equation}
    \mathcal{L}_{r} = \frac{1}{N}\bigg(\left\Vert\hat{\bm{\mathcal{F}}}+\hat{\textbf{K}}\odot \nabla{\hat{\textbf{h}}}\right\Vert_2^2+\left\Vert \nabla{\cdot \hat{\bm{\mathcal{F}}}}-\textbf{q}\right\Vert_2^2\bigg),
    \label{eq:residual loss}
\end{equation}
where $N$ is the number of image pixels (discretized grids), $\hat{\textbf{h}}$ represents the generated predictions, and $\nabla{\hat{\textbf{h}}}=[\hat{\textbf{h}}_x,\hat{\textbf{h}}_y]$, while the subscripts $x$ and $y$ represent calculating the image gradients along the horizontal and vertical axis using the Sobel filter. $\nabla{\cdot\bm{\mathcal{\hat{F}}}}=(\hat{\bm{\mathcal{F}}}_1)_x+(\hat{\bm{\mathcal{F}}}_2)_y$, in which $\hat{\bm{\mathcal{F}}}_1$ and $\hat{\bm{\mathcal{F}}}_2$ are the predicted flux field components along the horizontal and vertical directions, respectively; $\odot$ represents the element-wise product. The source term $\textbf{q}$ is set  as zero in this work.

Apart from the residual loss, we also enforce constraints to the boundaries, which can be defined as
\begin{equation}
    \mathcal{L}_{b} = \frac{1}{M}\bigg(\left\Vert \hat{\textbf{h}}(\textbf{x}_D)-\textbf{h}_D\right\Vert_2^2 + \left\Vert\hat{\bm{\mathcal{F}}}(\textbf{x}_N)-\bm{\mathcal{F}}_N\right\Vert_2^2\bigg),
    \hspace{0.5cm}
    \textbf{x}_D \in \Gamma_D,
    \hspace{0.1cm}
    \textbf{x}_N \in \Gamma_N,
    \label{eq:boundary loss}
\end{equation}
where $M$ is the number of pixels along all the boundaries, $\hat{\textbf{h}}(\textbf{x}_D)$ are generated predictions in Dirichlet boundaries $\Gamma_D$, and $\textbf{h}_D$ are known. Likewise, $\hat{\bm{\mathcal{F}}}(\textbf{x}_N)$ are generated predictions along the Neumann boundaries $\Gamma_N$, while $\bm{\mathcal{F}}_N$ are their corresponding reference values. Eventually, the incorporation of physics can be realized by adding these physical constraints to the original generator loss, i.e.
\begin{equation}
    \mathcal{L}_{\mathcal{G}}^{\rm new} = \mathcal{L}_\mathcal{G} + \lambda_r \mathcal{L}_r + \lambda_b \mathcal{L}_b,
    \label{eq:modified generator loss}
\end{equation}
in which $\mathcal{L}_\mathcal{G}$ is the original generator loss defined in equation (\ref{eq:GAN loss}), and $\lambda_r$ and $\lambda_b$ are coefficients for the additional physical constraints. Therefore, the physics-informed WGAN-GP can be trained by minimizing the discriminator loss (equation (\ref{eq:GAN loss})) and the new generator loss (equation (\ref{eq:modified generator loss})) iteratively. A schematic of how to train the physics-informed WGAN-GP is shown in Figure \ref{fig:physics-based WGAN-GP}, and the pre-trained model will be used in the subsequent semantic inpainting. For further details of training WGAN-GP, one can refer to the original paper \cite{gulrajani2017improved}.

\subsection*{3.2 Semantic inpainting via physics-informed WGAN-GP}
The task of semantic inpainting is to fill in missing regions of an image using the known pixels and a prior of what the images should look like. This is very similar to inferring the large-scale geological field based on some point measurements and prior information, like the spatially statistical property of the geological field. Deep generative models, like deep convolutional GANs \cite{radford2015unsupervised}, have been used recently for this task, and there exist some successful cases where semantic inpainting was applied to geostatistical modeling \cite{dupont2018generating,zhang2019generating,yeh2017semantic}. However, they only used direct point measurements to conduct semantic inpainting; in other words, only the hydraulic conductivity ($K$) point measurements were used  to infer the entire $K$ field. In this paper, we propose a physics-informed WGAN-GP framework with the aim of incorporating indirect measurements by enforcing the proper physics of the problem.

Upon achieving the trained physics-informed WGAN-GP, the generator $\mathcal{G}$ can produce diverse realizations of $\textbf{y}=[\textbf{K},\textbf{h}]$ that satisfy the equation (\ref{eq:steady-state flow}) by sampling $\textbf{z}\sim P_{\textbf{z}}$ and mapping $\textbf{y}=\mathcal{G}(\textbf{z})$. Based on the trained $\mathcal{G}$ and $\mathcal{D}$, we would like to generate realistic samples conditioned on a set of point measurements $\textbf{K}_{\rm obs}$ and $\textbf{h}_{\rm obs}$. This can be achieved by fixing the weights in $\mathcal{G}$ and $\mathcal{D}$, and optimizing $\textbf{z}$ to generate realistic samples based on the point measurements.

In order to guarantee the authenticity of generated samples, we would like the sample $\textbf{y}$ to closely support the real data distribution $P_{\rm data}$, that is to say, the generated sample should be assigned high scores by the discriminator $\mathcal{D}$. This can be realized by enforcing a prior loss that penalizes unrealistic samples. The prior loss is designed according to the original generator loss in equation (\ref{eq:GAN loss}), i.e.
\begin{equation}
    \mathcal{L}_{p}(\textbf{z}) = -\mathcal{D}(\mathcal{G}(\textbf{z})).
    \label{eq:prior loss}
\end{equation}
We would also like the generated samples to honor the point measurements, namely the generated predictions at measurement locations should totally match the observed values. This can be achieved by applying context loss, which is used to penalize the mismatch between the predictions and the measurements. By leveraging the masking matrix $\textbf{M}$, which has value 1 at the measurement locations and 0 otherwise, we can define the context loss as follows \cite{yeh2017semantic},
\begin{equation}
    \mathcal{L}_c(\textbf{z}|\textbf{y},\textbf{M}) = \left\Vert \textbf{M}_{\textbf{K}} \odot(\mathcal{G}(\textbf{z})^{\#\textbf{K}}-\textbf{K}_{\rm obs})\right\Vert_1 +\left\Vert \textbf{M}_{\textbf{h}} \odot(\mathcal{G}(\textbf{z})^{\#\textbf{h}}-\textbf{h}_{\rm obs})\right\Vert_1,
    \label{eq:context loss}
\end{equation}
where $\#\textbf{K}$ and $\#\textbf{h}$ represent their corresponding output channels in the predictions $\mathcal{G}(\textbf{z})$; $\textbf{M}_{\textbf{K}}$ and $\textbf{M}_{\textbf{h}}$ are corresponding masking matrices for $\textbf{K}$ and $\textbf{h}$. It is worth noting that the number of $\textbf{K}$ measurements is smaller than that of $\textbf{h}$. Thus, the total loss for finding the optimal latent variable $\textbf{z}$ is defined as
\begin{equation}
    \mathcal{L}_{\textbf{z}} = \mathcal{L}_c(\textbf{z}|\textbf{y},\textbf{M})+\lambda_p \mathcal{L}_p(\textbf{z}),
    \label{eq:total inpainting loss}
\end{equation}
where $\lambda_p$ is a coefficient that controls the trade-off between generating realistic samples and matching the point measurements.

\section*{4. Results}
\subsection*{4.1 Unconditional modeling by physics-informed WGAN-GP}
We consider steady-state flow in randomly heterogeneous porous media governed by equations (\ref{eq:steady-state flow}) and (\ref{eq:boundary conditions}). The domain is $\mathcal{X}=[0,2\,\rm{L}]\times[0,2\,\rm{L}]$, and the left and right boundaries are set as Dirichlet, with hydraulic head ($h$) values 1 $\rm{L}$ and 0 $\rm{L}$, respectively. The upper and lower boundaries are assigned Neumann with zero flux. The source field ($q$) is zero.

\begin{figure}
    \centering
    \subfigure[]{
    \includegraphics[width=0.9\textwidth]{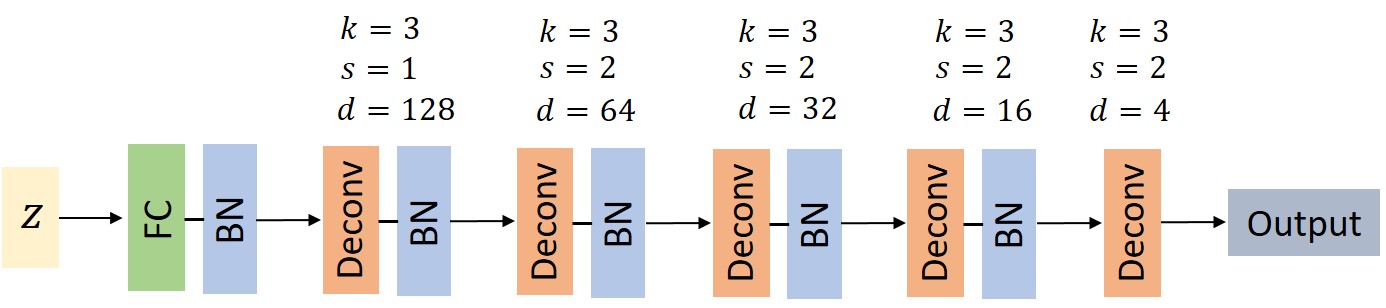}}
    \subfigure[]{
    \includegraphics[width=0.9\textwidth]{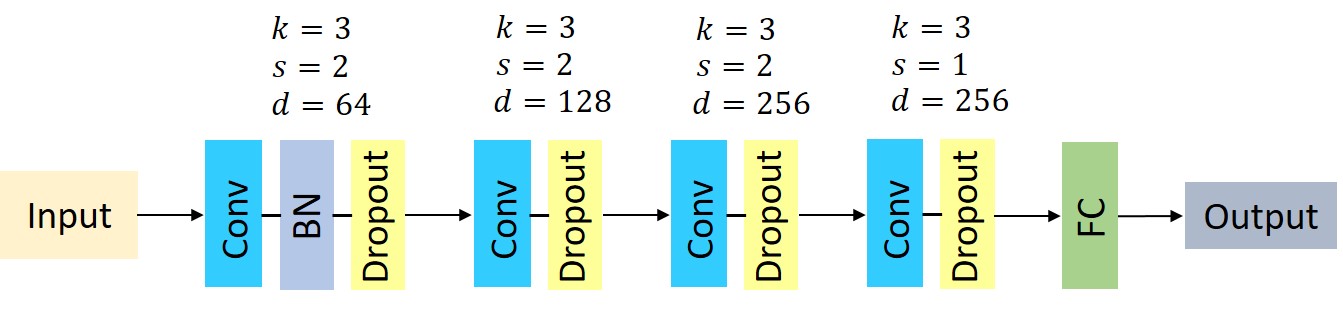}}
    \caption{The network architecture of (a) generator and (b) discriminator. In the generator network, the 100-dimensional random Gaussian noises serve as the input, going through a fully-connected layer (FC) and a set of batch normalization (BN) layers and deconvolutional layers. Also, $k$ represents the edge length of square kernel, $s$ is the stride size, and $d$ is the number of feature maps. In the discriminator network, the inputs are the images with four channels, which then go through a set of convolutional layers, batch normalization layers and dropout layers. The dropout rate is set as 0.3 in this work. The eventual fully-connected layer makes the outputs to be one-dimensional so that they can represent the scores of samples.}
    \label{fig:network architecture}
\end{figure}

The hydraulic conductivity $K$ is assumed to be a random field following a log-normal distribution. Specifically, it can be written as $K(\textbf{x})=\exp(G(\textbf{x}))$, where the vector $\textbf{x}=(x_1,x_2)$ represents two directions along the $x-$ and $y-$ axes, respectively. $G(\textbf{x})$ is a Gaussian process defined as $G(\cdot)\sim \mathcal{GP}(\mu,k(\cdot,\cdot))$, where $k(\textbf{x},\textbf{x}')=\sigma^2 \exp[-(  {(x_1-x_1')^2}/{2l_1^2} + {(x_2-x_2')^2}/{2l_2^2} )]$, $\mu$ and $\sigma^2$ are mean and variance of ${\rm ln}K$, and $l_1$ and $l_2$ are the correlation lengths along the two axes. In this case, we set $\mu=0$, $\sigma^2=1$, and $l_1=l_2=0.5$ $\rm{L}$. The ${\rm ln}K$ field realizations are discretized over $64\times64$ grids, and generated by a Karhunen-Lo\`eve expansion with leading 512 terms retained, which corresponds to 95.04\% energy. The hydraulic head $h$, horizontal flux $\mathcal{F}_1$, and vertical flux $\mathcal{F}_2$ with respect to each ${\rm ln}K$ realization are calculated by the numerical solver ModFlow \cite{harbaugh2005modflow}. The training datasets are composed of these data in the form of multi-channel images. To be specific, as shown in Figure~\ref{fig:physics-based WGAN-GP}, we put ${\rm ln}K$ in the first channel, $h$ in the second channel, and $\mathcal{F}_1$ and $\mathcal{F}_2$ in the third and fourth channels, respectively. The Adam optimizer~\cite{kingma2014adam} with learning rate $1e-4$ is utilized to train the $\mathcal{G}$ and $\mathcal{D}$ networks, whose architectures are shown in Figure \ref{fig:network architecture}. The coefficients for the physical constraints are selected as $\lambda_r=1$ and $\lambda_b=10$. The training data are composed of 10,000 samples, and we train the networks for 150,000 iterations with batch size 50, which requires about 10 hours running on one GPU (Telsa V100). Additionally, each iteration for training $\mathcal{G}$ corresponds to five iterations for training $\mathcal{D}$, which follows the basic training scheme of the original WGAN-GP \cite{gulrajani2017improved}.

\begin{figure}
    \centering
    \subfigure[]{
    \includegraphics[width=0.45\textwidth]{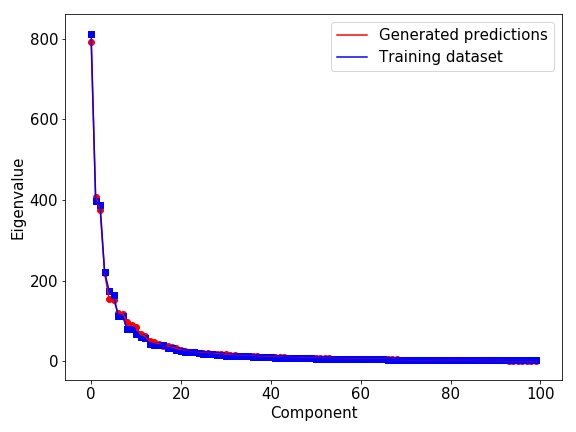}}
    \subfigure[]{
    \includegraphics[width=0.45\textwidth]{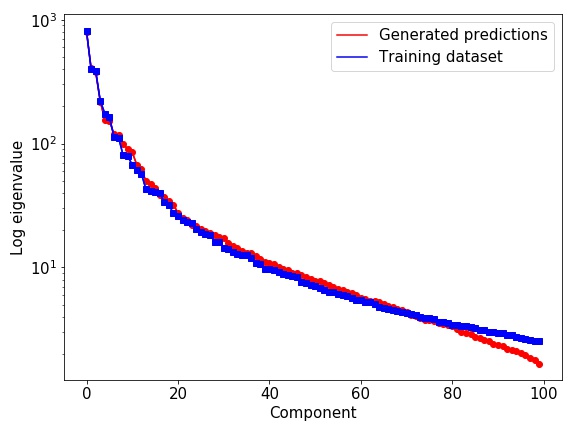}}
    \subfigure[]{
    \includegraphics[width=0.45\textwidth]{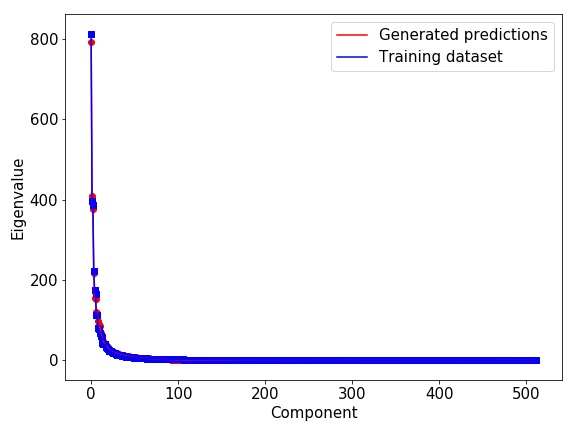}}
    \subfigure[]{
    \includegraphics[width=0.45\textwidth]{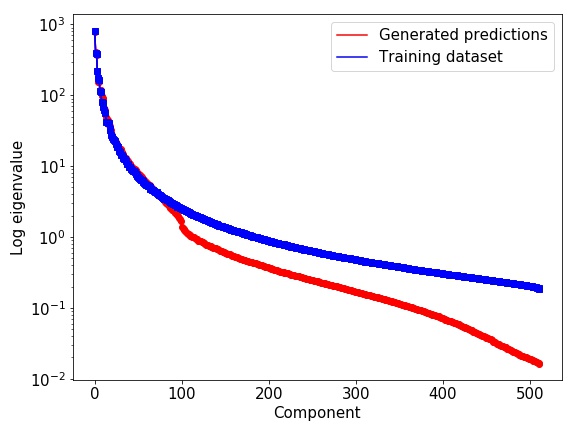}}
    \caption{(a) The spectrum of ${\rm ln}K$ training datasets and generated predictions with 100 largest eigenvalues. The top 100 eigenvalues of the training datasets account for 93.5\% of the total energy, while that of the generated predictions
    accounts for 92.8\%. (b) Plot of log-transformed values of (a); eigenvalues after the 80th are different. (c) Plot showing the spectral performance with 512 largest eigenvalues, and (d) Corresponding plot of its log transformation. The 512 largest eigenvaues account for 100\% of total energy, while 97.4\% for the
    generated predictions.}
    \label{fig:spectra matching of lnk}
\end{figure}

\begin{figure}
    \centering
    \subfigure[]{
    \includegraphics[width=0.45\textwidth]{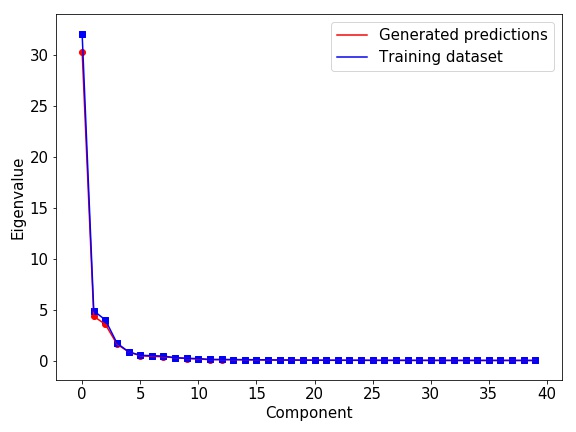}}
    \subfigure[]{
    \includegraphics[width=0.45\textwidth]{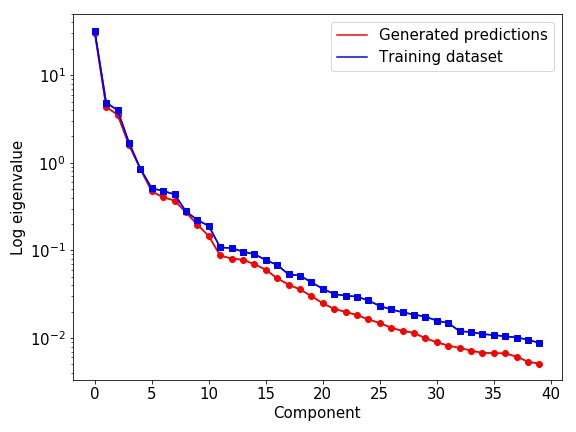}}
    \caption{(a) The spectral matching performance of the $h$ training dataset and generated predictions, and (b) Plot showing the corresponding log-transformed values. Because the pattern of $h$ is much simpler than ${\rm ln}K$, only 40 eigenvalues are plotted here. The largest 40 eigenvalues of the training datasets account for 99.9\% of total energy, while that of generated images account for 92.0\%.}
    \label{fig:spectra matching of h}
\end{figure}

With the trained model, we can generate new realizations by sampling $\textbf{z}\sim P_{\textbf{z}}$ and feeding it to $\mathcal{G}$. Here, we sample $\textbf{z}$ for 10,000 times, with the same size as training datasets, and generate the corresponding outputs ${\rm ln}K$, $h$, $\mathcal{F}_1$, and $\mathcal{F}_2$ distributed in four channels, respectively. The correlation structure spectrum of the random field is used as a metric to measure the performance of unconditional generated predictions, i.e., without any point measurements. As shown in Figure \ref{fig:spectra matching of lnk}, \ref{fig:spectra matching of lnk}a is the spectra of ${\rm ln}K$ training datasets versus that of generated predictions with 100 largest eigenvalues. The top 100 eigenvalues of training datasets account for 93.5\% energy, while that of generated predictions account for 92.8\%. To better demonstrate the differences in the components with relatively smaller eigenvalues, we make the log-transformation of Figure \ref{fig:spectra matching of lnk}a to \ref{fig:spectra matching of lnk}b, and we observe a deviation after the 80th eigenvalue. Since the ${\rm ln}K$ field is generated by retaining 512 leading KL terms, by contrast, we conduct the eigenvalue decomposition of ${\rm ln}K$ predictions with 512 terms kept. We see in Figures \ref{fig:spectra matching of lnk}c and \ref{fig:spectra matching of lnk}d, that the log-eigenvalues after the 100th are different, but this part contributes insignificantly to overall energy (less than 2.6\%). Likewise, we compare the spectra of $h$ field given by the training datasets and generated predictions. Since the correlation structure of $h$ field is much simpler than that of ${\rm ln}K$ field, we only compare the spectra with the 40 leading eigenvalues retained. The largest 40 eigenvalues of the training datasets account for 99.9\% energy, while those of the generated predictions account for about 92.0\%. As shown in Figure $\ref{fig:spectra matching of h}$, the first component and the log-components after the 10th exhibit gaps between the training datasets and generated predictions. Therefore, with respect to spectra matching, we believe that the WGAN-GP model has learned the spatial correlations from the training datasets and is able to generate the samples with very similar spatial structures as the training datasets.


\begin{figure}
    \centering
    \includegraphics[width=\textwidth]{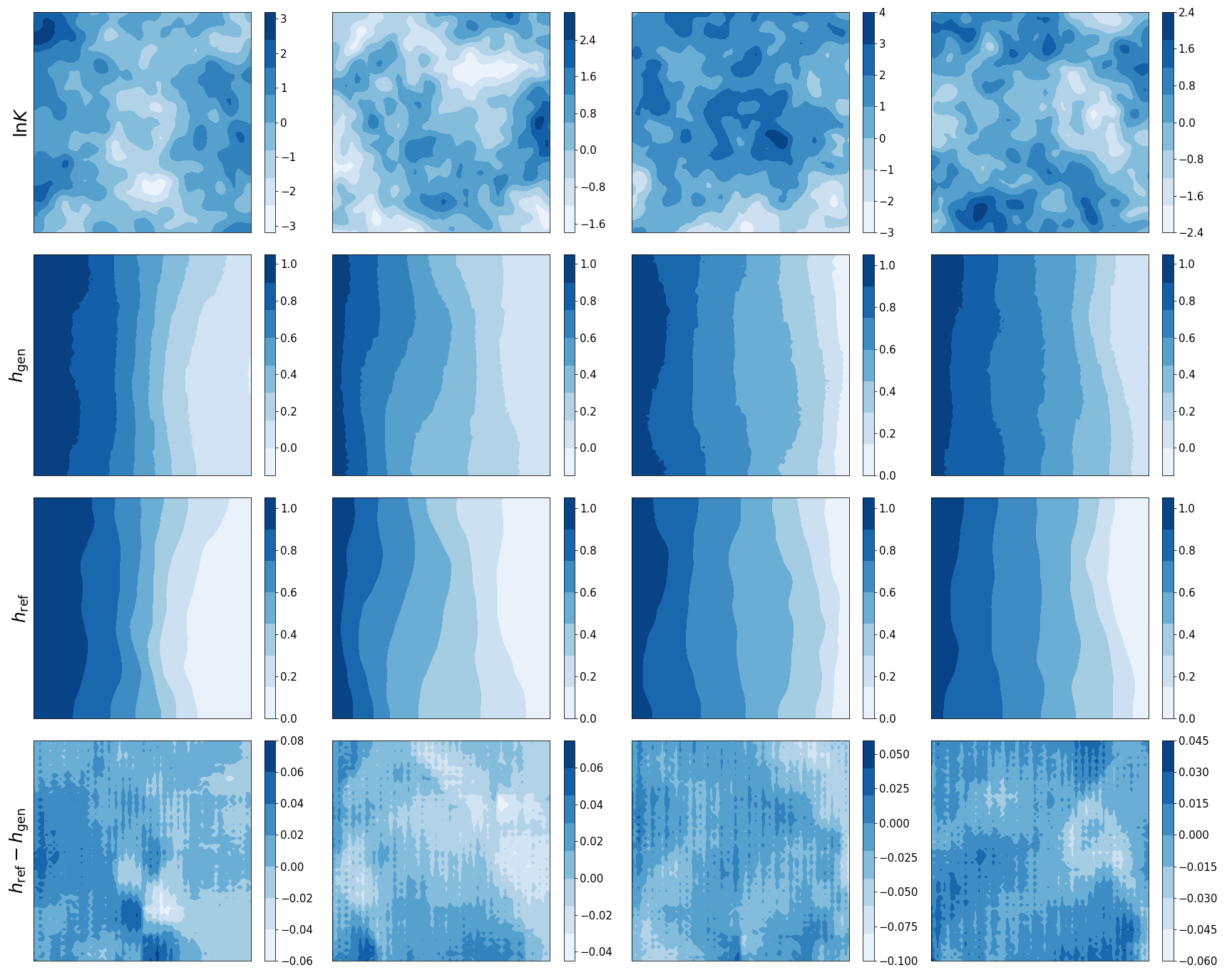}
    \caption{The unconditional realizations of log conductivity (${\rm ln}K$, see the first row) and pressure heads ($h$, see the second row) given by the trained WGAN-GP. The third row shows the corresponding numerical solutions by feeding the ${\rm ln}K$ to ModFlow, and the respective differences between the four numerical solutions and generated predictions are shown in the last row. The ${\rm ln}K$ realizations are both realistic and diverse. The $h$ realizations provided by WGAN-GP and ModFlow are quite similar (with very small differences), even though the results of WGAN-GP are somewhat noisy, which may attribute to the checkerboard effects in the transposed convolutional processes \cite{odena2016deconvolution}.}
    \label{fig:unconditional realizations of para and state}
\end{figure}

\begin{figure}
    \centering
    \subfigure[]{
    \includegraphics[width=0.43\textwidth]{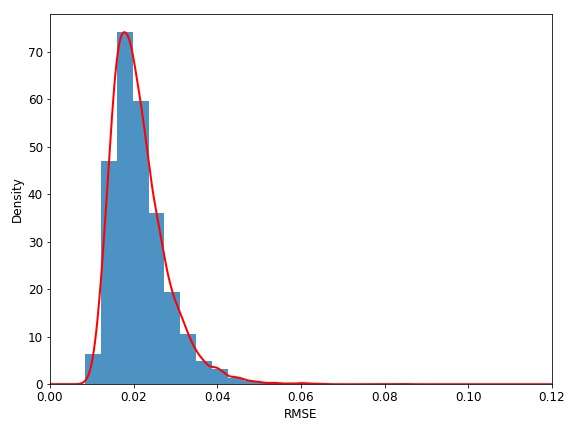}}
    \hspace{0.5cm}
    \subfigure[]{
    \includegraphics[width=0.43\textwidth]{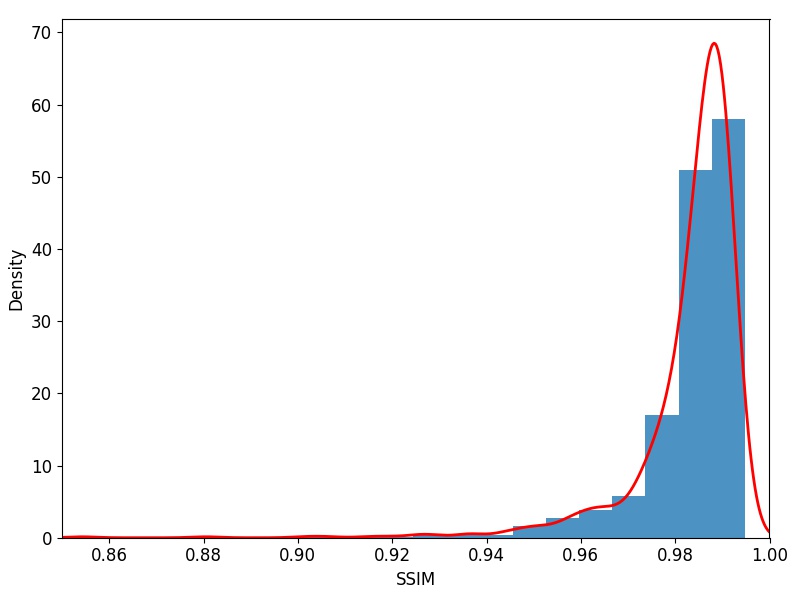}}
    \caption{(a) Root mean square error (RMSE), and (b) structural similarity index (SSIM) of $h$ generated by WGAN-GP versus the ones being generated by the numerical solver ModFlow, which is fed with the ${\rm ln}K$ as parameters generated by WGAN-GP. The RMSE values are very close to 0 and SSIM values are close to 1. This means that the correct physics has been incorporated into  WGAN-GP and can be embedded in the predictions of different channels. The mean RMSE is 0.02, and the mean SSIM is 0.99.}
    \label{fig: rmse and ssim}
\end{figure}

To further verify whether the physical information (equation (\ref{eq:steady-state flow})) has been incorporated in the model, we calculate the corresponding $h$ solutions by using ModFlow, which takes the $K$ predictions as input parameters. The boundary conditions are the same as those in building the training datasets. We randomly select four samples to demonstrate the performance. As shown in Figure \ref{fig:unconditional realizations of para and state}, there are four generated realizations of ${\rm ln}K$ (see the first row) and corresponding $h$ predictions (see the second row). The third row in Figure \ref{fig:unconditional realizations of para and state} represents the corresponding numerical solutions of ${\rm ln}K$ given by ModFlow, and the last row shows the respective differences between the numerical solutions and generated predictions. We observe that the patterns of WGAN-GP predictions are very similar with those of numerical solutions, even though the latter is smoother than the former. To measure the accuracy, the commonly used root mean sqaure error (RMSE) and the structural similarity index (SSIM)~\cite{wang2004image} widely used in image analysis are adopted here as evaluating metrics, which are defined by
\begin{equation}
    {\rm RMSE(\textbf{u},\textbf{v})}=\sqrt{\frac{1}{N}\left\Vert \textbf{u}-\textbf{v}\right\Vert_2^2},
    \label{eq: rmse}
\end{equation}

\begin{equation}
    {\rm SSIM(\textbf{u},\textbf{v})} = \frac{(2\mu_{\textbf{u}}\mu_{\textbf{v}}+C_1)(2\sigma_{\textbf{uv}}+C_2)}{(\mu_{\textbf{u}}^2+\mu_{\textbf{v}}^2+C_1)(\sigma_{\textbf{u}}^2+\sigma_{\textbf{v}}^2+C_2)},
    \label{eq:ssim}
\end{equation}
where $\textbf{u}$ and $\textbf{v}$ are two image signals, $N$ is the number of pixels of each image, $\mu$ and $\sigma$ represents the mean and standard deviation, respectively. $C_1$ (0.01) and $C_2$ (0.03) are small constants used to stabilize the calculation. The range of SSIM is from -1 to 1, and the closer to 1, the faster is the convergence for the two images to be identical. We compute two metrics among 10,000 samples and plot their distributions in Figure \ref{fig: rmse and ssim}. It is obvious that the vast majority of RMSE values are around 0.02 and the SSIM values are mostly larger than 0.98, which means the numerical solutions are quite similar with the WGAN-GP predictions, and hence the physical information has been incorporated into the WGAN-GP model.

\subsection*{4.2 Inpainting using trained physics-informed WGAN-GP}
Having verified that the physics is correctly incorporated into the trained WGAN-GP, we can proceed with the semantic inpainting, that is, generating realizations that honor the point measurements. As introduced in the section 3.2, the goal of inpainting is to find the optimal latent variable $\textbf{z}$ underlying the point measurements, while taking the authenticity of generated predictions into consideration simultaneously by adjusting the trade-off coefficient, i.e., $\lambda_p=0.1$ in this work. It should be noted that the preset dimension of $\textbf{z}$ has an effect in the inpainting performance, and we justify why we choose $\textbf{z}_{\rm dim}=100$, see the \ref{appendix}. The Adam optimizer~\cite{kingma2014adam} with learning rate $1e-2$ is utilized to find the optimal $\textbf{z}$. However, the loss function in equation (\ref{eq:total inpainting loss}) is non-convex, and thus the obtained $\textbf{z}$ is a local optimal value. Therefore, we run each scenario for 10 times to alleviate the effects of randomness.

\begin{figure}
    \centering
    \includegraphics[width=\textwidth]{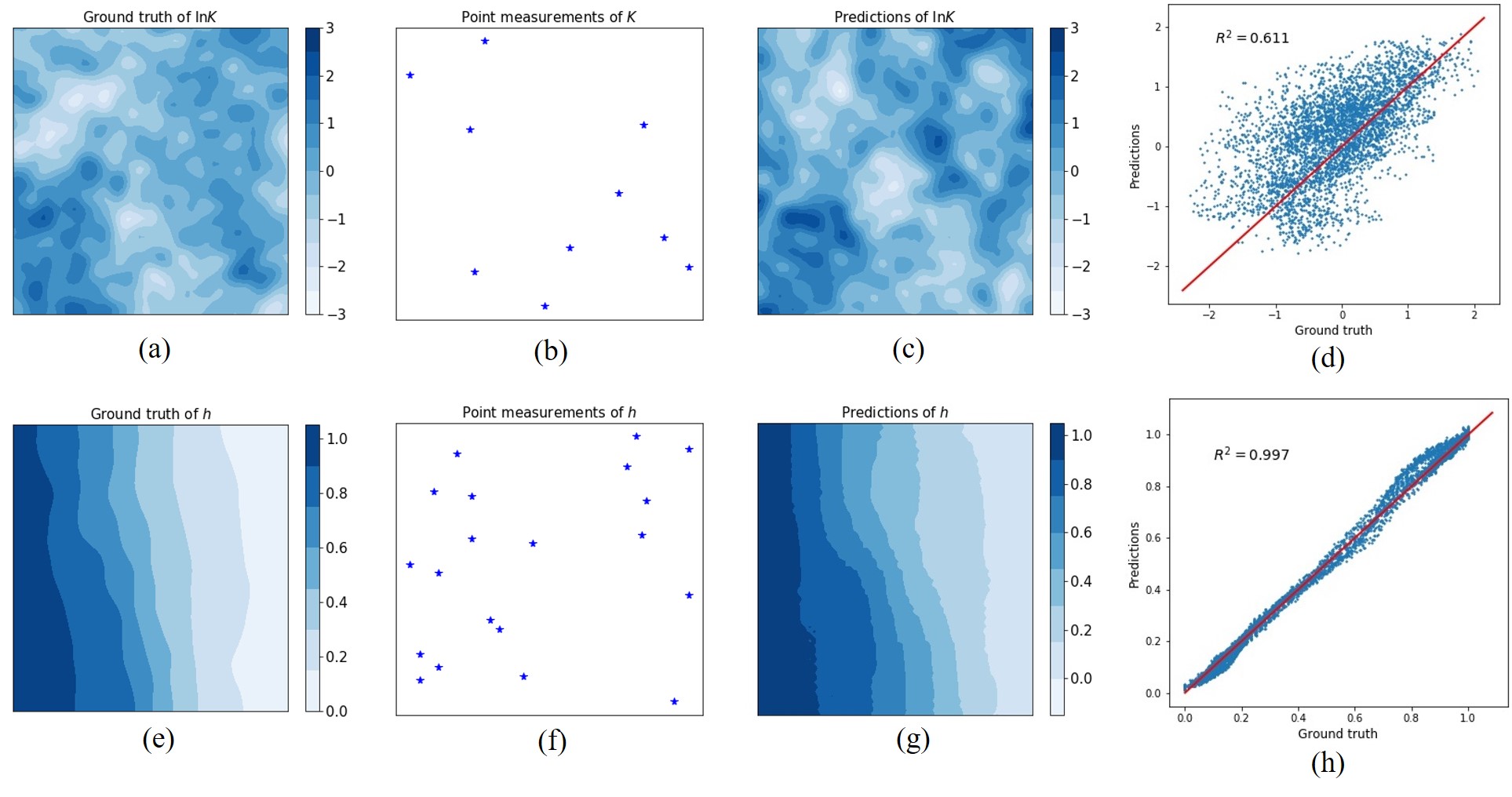}
    \caption{The inpainting results of ${\rm ln}K$ and $h$ field. (a) and (e) are ground truths of ${\rm ln}K$ and $h$, respectively. We randomly select some points from the respective ground truth to serve as the measurements; here (b) $N_K=10$, and (f) $N_h=20$. (c) and (g) are conditional generated predictions that honor their respective measurements. The scatter plot of ${\rm ln}K$ ground truth versus its predictions is shown in (d), while $h$ field is shown in (h). To alleviate the effects of randomness, we test the case for 10 times, and the predictions shown here are the mean values. The scatter plots are also mean predictions versus the ground truth. The scatter plot of ${\rm ln}K$ in (d) reflects the complexity of its pattern compared to the smoother spatial patterns of $h$ in figure (h).}
    \label{fig:conditional_case_1}
\end{figure}

\begin{figure}
    \centering
    \includegraphics[width=\textwidth]{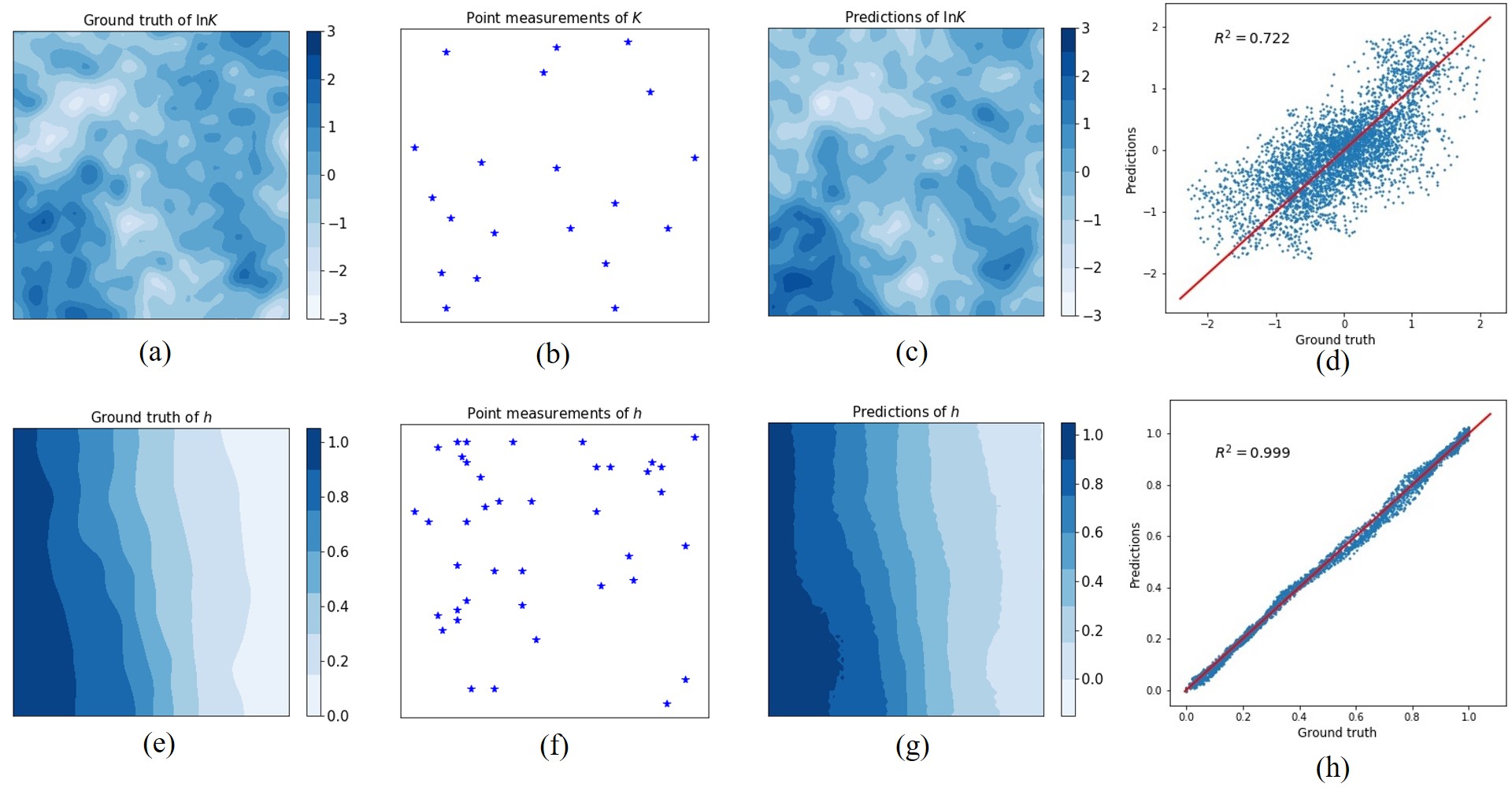}
    \caption{The inpainting results of ${\rm ln}K$ and $h$ field. Based on the same ground truth shown in Figure \ref{fig:conditional_case_1} (a) and (e), we add point measurements to $N_K=20$, $N_h=40$, which are shown in (b) and (f), respectively. As the same layout of Figure \ref{fig:conditional_case_1}, (c) and (g) are respective conditional realizations, and (d) and (h) are scatter plots of ground truth versus their predictions. Likewise, the prediction results here are also the average of running 10 times.}
    \label{fig:conditional_case_2}
\end{figure}

As shown in Figures \ref{fig:conditional_case_1}a and \ref{fig:conditional_case_1}e, we prepare a paired ground truth of ${\rm ln}K$ and $h$ that satisfy the equations (\ref{eq:steady-state flow}) and (\ref{eq:boundary conditions}), and randomly select 10 $K$ points ($N_K=10$) and 20 $h$ ($N_h=20$) points to be known and serve as the measurements (Figures \ref{fig:conditional_case_1}b and \ref{fig:conditional_case_1}f). Figures \ref{fig:conditional_case_1}c and \ref{fig:conditional_case_1}g are the conditional generated predictions (means of 10 runs) given by trained $\mathcal{G}$ while honoring their respective point measurements. Obviously, the pattern of reconstructed $h$ field is much more similar to the ground truth than that of ${\rm ln}K$ field, because the $h$ field has much simpler spatial structure, which is also demonstrated in the spectral performance shown in Figures \ref{fig:spectra matching of lnk} and \ref{fig:spectra matching of h}. The scatter plots of the ground truth versus its predictions, shown in Figures \ref{fig:conditional_case_1}d and \ref{fig:conditional_case_1}h, clearly demonstrate their matching performance. Based on the same ground truth, we add the point measurements to $N_K=20$ and $N_h=40$, as shown in Figures \ref{fig:conditional_case_2}b and \ref{fig:conditional_case_2}f. Their respective conditional generated predictions are given in Figures \ref{fig:conditional_case_2}c and \ref{fig:conditional_case_2}g. We see that the pattern of reconstructed ${\rm ln}K$ becomes more similar to the ground truth compared to the last case (see Figure \ref{fig:conditional_case_1}c), which is also confirmed by the scatter plot (Figure \ref{fig:conditional_case_2}d) with larger $R^2$ (coefficient of determination) value than the previous one (Figure \ref{fig:conditional_case_1}d).

\begin{figure}
    \centering
    \subfigure[]{
    \includegraphics[width=0.45\textwidth]{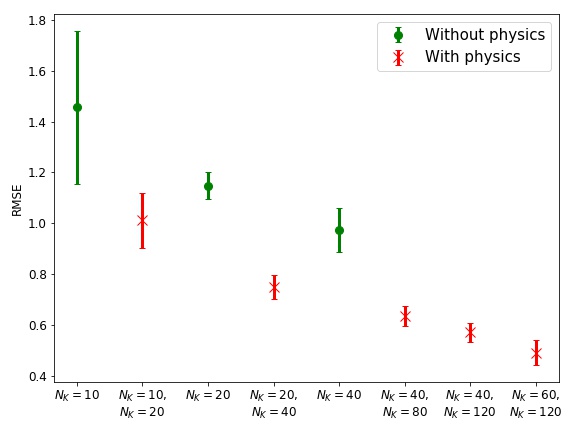}}
    \hspace{0.2cm}
    \subfigure[]{
    \includegraphics[width=0.45\textwidth]{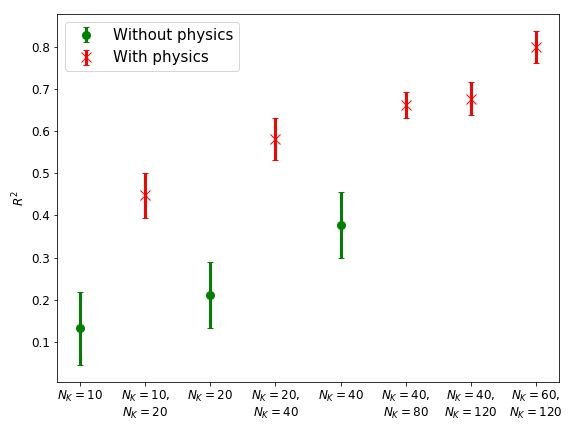}}
    \caption{Effect of incorporating physics into semantic inpainting. (a) Root mean square error (RMSE), and (b) coefficient of determination ($R^2$) between the ${\rm ln}K$ ground truth and its conditional generated predictions in different cases. To alleviate randomness, we run each case for 10 times, and plot the mean and standard deviations of the two metrics. In Case 1 ($N_K=10$), Case 3 ($N_K=20$) and Case 5 ($N_K=40$), only direct measurements are used for inpainting of ${\rm ln}K$ field. Based on the three cases, we add indirect measurements and enforce the physical constraints, i.e. Case 2 ($N_K=10,N_h=20$), Case 4 ($N_K=20,N_h=40$) and Case 6 ($N_K=40,N_h=80$), and both of the two metrics show improvement. Then, we continue to add indirect measurements (Case 7: $N_K=40,N_h=120$) based on Case 6, and add direct measurements (Case 8: $N_K=60,N_h=120$) based on Case 7. The results show that adding direct measurements make larger contributions to the optimality of the two metrics, which is also in line with the intuition.}
    \label{fig:RMSE and R2}
\end{figure}

In this work, we improve the inpainting of ${\rm ln}K$ field by utilizing the physical connection between  $h$  and ${\rm ln}K$. To illustrate this, we consider eight scenarios here and adopt the commonly used RMSE and $R^2$ as evaluating metrics. Considering the effects of local optimal, we run each case for 10 times based on the same measurements settings and calculate the means and standard deviations of two metrics. As shown in Figure \ref{fig:RMSE and R2}, the labels in $x-$ axis from left to right represent the number of point measurements in cases from scenario 1 to 8 in sequence. In Case 1 ($N_K=10$), Case 3 ($N_K=20$) and Case 5 ($N_K=40$), only direct measurements are used to reconstruct the ${\rm ln}K$ field. Based on the three cases, indirect measurements ($h$) are incorporated in the Case 2 ($N_K=10$, $N_h=20$), Case 4 ($N_K=20$, $N_h=40$) and Case 6 ($N_K=40$, $N_h=80$). It is obvious that both the RMSE and $R^2$ can achieve better performance in the cases where the increased number of indirect $h$ measurements are fused. To test the effect of measurement type, we consider two further cases. Based on Case 6 ($N_K=40$, $N_h=80$), we add indirect measurements to form Case 7 ($N_K=40$, $N_h=120$), and subsequently add direct measurements to make up Case 8 ($N_K=60$, $N_h=120$) based on Case 7. We can see from the Figure \ref{fig:RMSE and R2} that adding direct ${\rm ln}K$ measurements can indeed make larger improvements with respect to two metrics, which is also in line with the intuition that direct measurements are more valuable, but of course at a higher cost.

\section*{5. Conclusion}
In this work, we extended the original semantic inpainting to a physics-informed version with an emphasis on jointly incorporating the direct and indirect measurements by exploiting the underlying physics. Specifically, we proposed a physics-informed WGAN-GP method which is pre-trained for learning the physical laws from a large number of unconditional realizations, and subsequently we employed it to perform the inpainting task. We used convolutional architectures to tackle high-dimensional images and the Sobel filter to encode physical constraints based on the governing mathematical model. We tested the proposed method in a groundwater flow problem, in which the heterogeneous hydraulic conductivity field ($K$) needs to be inferred from some direct measurements and indirect pressure head ($h$) measurements. Our results demonstrated that the physical information can be effectively incorporated into the pre-trained physics-informed WGAN-GP and the $h$ measurements can indeed help improve the inpainting results of the $K$ field.

The findings of our study have important implications for tackling high-dimensional problems by leveraging deep learning models, while simultaneously satisfying the underlying physical laws, the prior spatial statistics, and the in situ measurements. An important open issue in designing effective experimental strategies to obtain the most informative measurements is to incorporate uncertainty quantification through active learning techniques, e.g., see preliminary results in \cite{zhang2019quantifying}. We plan to endow the proposed physics-informed WGAN-GP with total uncertainty quantification in future work.

\section*{Acknowledgments}
The first author acknowledges the support from the National Key Research and Development Program of China (2018YFC1800501) and China Scholarship Council scholarship (201806320117). The last author acknowledges the support from the PhILMs grant DE-SC001954.


\begin{appendix}
\section{Testing the effects of input noise dimension on inpainting results}
\label{appendix}

In the physics-informed WGAN-GP framework, the dimension of input noise ($\textbf{z}$)  will influence how much spatial variability can be captured by the model, however, it is usually determined artificially without following a rigorous approch. Based on a pre-trained model, the dimension of $\textbf{z}$ is fixed, and this may have an effect on the performance of semantic inpainting. To test this effect, we consider five cases here, in which only $\textbf{z}$ dimension is different when training the physics-informed WGAN-GP, and all the other settings as the same as that in section 4.1. The dimension settings and the corresponding retained energy are shown in Table \ref{tab:long}. After obtaining the trained model with different input dimensions, we conduct semantic inpainting subsequently. It is worth pointing that the measurement settings are the same among the five cases, i.e., $N_K=20$, $N_h=40$, which are randomly selected from a ground truth. The root mean square error (RMSE) and coefficient of determination ($R^2$) serve as the evaluating metrics, and we run each case for 10 times to alleviate the randomness.

\begin{figure}
    \centering
    \subfigure[]{
    \includegraphics[width=0.45\textwidth]{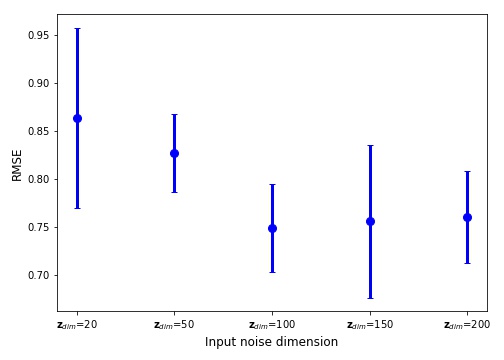}}
    \hspace{0.2cm}
    \subfigure[]{
    \includegraphics[width=0.45\textwidth]{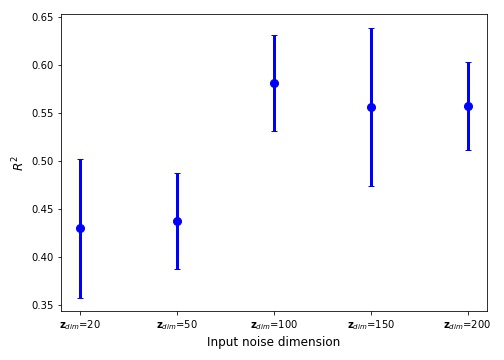}}
    \caption{The (a) root mean square error (RMSE) and (b) coefficient of determination ($R^2$) between the ${\rm ln}K$ ground truth and its generated predictions. From Case A1 ($\textbf{z}_{\rm dim}=20$) to  Case A3 ($\textbf{z}_{\rm dim}=100$), the two metrics both get better. However, when increasing $\textbf{z}_{\rm dim}$ further, the results do not show improvement, and in fact $R^2$ deteriorates.}
    \label{fig:rmse and r2 of inpainting}
\end{figure}

We see in Figure \ref{fig:rmse and r2 of inpainting} that the input noise dimension does have some effect on the inpainitng results. From Case A1 ($\textbf{z}_{\rm dim}=20$) to Case A3 ($\textbf{z}_{\rm dim}=100$), both of the metrics become better remarkably, which we can attribute to the fact that a higher input noise dimension is more beneficial to capturing the spatial variability. However, by increasing the dimension further, we cannot witness a positive improvement of the two metrics, and in fact $R^2$ deteriorates. We think the reason behind is that, in the inpainting process, the optimal latent variable with high dimension is relatively more difficult to be found compared to that with low dimension. In this test, the point measurements, i.e, $N_K=20$, $N_h=40$, may not provide enough information to determine the underlying optimal latent variable whose dimension is larger than 150. Therefore, in terms of the dimension of input noise, there is a trade-off between the capturing energy and finding optimal value in the inpainting task. Consequently, we make a relatively safe choice, i.e., $\textbf{z}_{\rm dim}=100$, in the all examples of this work.

\begin{center}
\begin{longtable}{|c|c|c|}
\caption{The dimension settings of input noise and their corresponding retained energy in five cases.}
\vspace{-0.3cm}
\label{tab:long} \\
 \hline
  Cases & Dimension & Retained energy \\
 \hline
 Case A1 & 20 & 76.8\% \\
 Case A2 & 50 & 87.4\% \\
 Case A3 & 100 & 93.5\% \\
 Case A4 & 150 & 95.2\% \\
 Case A5 & 200 & 96.6\% \\
 \hline
\end{longtable}
\end{center}

\end{appendix}

\bibliographystyle{elsarticle-num}
\biboptions{sort&compress}
\bibliography{main}







\end{document}